\documentclass[sigconf]{acmart}

\usepackage{booktabs} 
\usepackage{hyperref}
\usepackage[capitalise,noabbrev,nameinlink]{cleveref}
\usepackage{longtable}
\usepackage{listings}

\usepackage{dcolumn} 

\usepackage{tikz} 
\usetikzlibrary{shapes,arrows,patterns} 
\usepackage{pgfplots} 
\usepgfplotslibrary{statistics}
\pgfplotsset{compat=1.14}  

\usepackage{mycommands}
\usepackage{myconfig}

\setcopyright{acmlicensed}

\acmDOI{10.1145/3297280.3297465}

\acmISBN{978-1-4503-5933-7/19/04}

\acmConference[SAC '19]{The 34th ACM/SIGAPP Symposium on Applied Computing}{April 8--12, 2019}{Limassol, Cyprus}
\acmYear{2019}
\copyrightyear{2019}

\acmArticle{4}
\acmPrice{15.00}

\acmBooktitle{The 34th ACM/SIGAPP Symposium on Applied Computing (SAC '19), April 8--12, 2019, Limassol, Cyprus}

\begin{document}
\title{A Real-Time Remote IDS Testbed for Connected Vehicles}

\author{Valentin Zieglmeier}
\orcid{0000-0002-3770-0321}
\affiliation{%
  \institution{Technical University of Munich}
  \city{Munich}
  \state{Germany}
}
\email{v.zieglmeier@tum.de}

\author{Severin Kacianka}
\orcid{0000-0002-2546-3031}
\affiliation{%
  \institution{Technical University of Munich}
  \city{Munich}
  \state{Germany}
}
\email{severin.kacianka@tum.de}

\author{Thomas Hutzelmann}
\orcid{0000-0002-2871-3905}
\affiliation{%
	\institution{Technical University of Munich}
	\city{Munich}
	\state{Germany}
}
\email{t.hutzelmann@tum.de}

\author{Alexander Pretschner}
\affiliation{%
	\institution{Technical University of Munich}
	\city{Munich}
	\state{Germany}
}
\email{alexander.pretschner@tum.de}

\renewcommand{\shortauthors}{Valentin Zieglmeier et al.}

\begin{abstract}
Connected vehicles are becoming commonplace. A constant connection between vehicles and a central server enables new features and services. This added connectivity raises the likelihood of exposure to attackers and risks unauthorized access.

A possible countermeasure to this issue are intrusion detection systems (IDS), which aim at detecting these intrusions during or after their occurrence. The problem with IDS is the large variety of possible approaches with no sensible option for comparing them.

Our contribution to this problem comprises the conceptualization and implementation of a testbed for an automotive real-world scenario. That amounts to a server-side IDS detecting intrusions into vehicles remotely. To verify the validity of our approach, we evaluate the testbed from multiple perspectives, including its fitness for purpose and the quality of the data it generates.

Our evaluation shows that the testbed makes the effective assessment of various IDS possible. It solves multiple problems of existing approaches, including class imbalance. Additionally, it enables reproducibility and generating data of varying detection difficulties. This allows for comprehensive evaluation of real-time, remote IDS.
\end{abstract}

%
%



\maketitle{}


\pgfplotsset{
	scalabilityplot/.style={
		mylines, footnoteaxislabels,
		width=0.7\linewidth, height=40mm,
		axis y line*=left,
		axis x line*=bottom,
		xlabel=\textit{Number of components},
		scatter/classes={a={mark=x,draw=black,mark size=2}}
	}
}
\pgfplotsset{
	errorbarplotstyle/.style={
		mark=*,black,mark size=1,only marks,
		mark options={fill=black,draw=black},
		error bars/.cd,
		y dir=both,y explicit,
		error bar style={draw=black,very thin}
	}
}

\pgfplotsset{
	reproducibilityplot/.style={
		height=40mm,
		legend style={at={(0.5,1.3)}},
		xlabel=\empty,
		ylabel=\textit{Probability density},
		cycle list={
			{fill=none,densely dashed,draw=TumDarkGrey},
			{fill=none,densely dashdotted,draw=TumMedGrey},
			{fill=none,densely dotted,draw=black},
			{fill=none,TumMedBlue},
			{fill=none,TumAccentGreen}
		}
	}
}

\pgfplotsset{
	histogramline/.style={
		legend style={at={(0.5,1.3)}},
		height=35mm,
		xtick=data,
		xmajorgrids,
		major x grid style={draw=TumMedGrey},
	}
}

\section{Introduction}

In 2015, 35~\% of new cars sold were already connected to the
Internet~\cite{accenture2015share}. Accenture estimates that by 2020 that number
will rise to 98~\%, reaching 100~\% in 2025~\cite{accenture2015share}.
Manufacturers push for higher connectivity, as it offers advantages such as the
ability to provide automated software updates which do not require customers to
visit a repair shop for software updates~\cite{brooks2009automobile}.
Additionally, new features enabled by Internet connections are used as selling
points. For example, vehicles may be monitored and controlled with a
smartphone~\cite{brooks2009automobile}. Finally, vehicle-to-vehicle
communication allows for advanced autonomous features such as cooperative
collision warnings~\cite{yang2004vehicle}. However, this connectivity also
raises the likelihood of exposing vulnerabilities, which increases the risk of
attacks~\cite{parkinson2017cyber}. \Citeauthor{checkoway2011comprehensive} argue
that the vehicle remote telematic systems providing a constant connection over
cellular networks are one of the most important parts of the wireless attack
surface~\cite{checkoway2011comprehensive}.

For passenger cars, traditional defense measures may not always be feasible. The vehicles are expected to subsist for long periods of time, with the average age of cars in the United States rising from $10.6$ years in 2010 to $11.6$ years in 2016~\cite{statista2016age}. It is reasonable to expect that it will rise even further in the future. \Citeauthor{zhang2014defending} argue that this long lifespan makes it hard for manufacturers to predict the necessary hardware for on-board protection~\cite{zhang2014defending}. Because of strict limits in production budgets, manufacturers are likely to minimize the cost for each vehicle and only include the minimum required security hardware. Therefore, off-board protection seems promising to provide sufficient security over the whole lifespan of the vehicle~\cite{hutzelmann2018comprehensive}. An important challenge with this approach is balancing on-board processing load and the communication of the vehicle to the manufacturer server~\cite{zhang2014defending}. Ideally, the existing vehicle communication is utilized for this purpose, as it would neither require additional processing nor communication.

\subsection{Definition}
\label{subsection:introduction-definition}

We define intrusions as deviations from the expected behavior of a system without the manufacturer's knowledge. On the one hand, this comprises malware, unauthorized access and anomalies, including software bugs, that mostly occur without the user's knowledge. On the other hand, our definition encompasses misuse and fraud that can take place with or without the user's knowledge. An example of intentional fraud by users are off-limits modifications intended to activate features that would otherwise be paid upgrades. Different types of intrusions are often difficult or impossible to differentiate from each other and do not always correspond to illegitimate actions. Our definition follows the NIST~\cite{scarfone2007guide}.

\subsection{Motivation and Requirements}

To defend against intrusions, one can employ intrusion detection systems (IDSs). They can be used as a second line of defense after preventive protection mechanisms such as software integrity verification~\cite{oguma2008new}. IDSs are built to detect intrusions that have already taken place, aiming to mitigate their consequences. For this purpose, they monitor events for possible violations of some defined policy or expected behavior~\cite{scarfone2007guide}. A considerable advantage of IDSs in the scenario of connected vehicles is that they can be employed non-invasively. That means that a complete off-board protection mechanism can be implemented with their help, solving the problems that we discussed before. Additionally, IDSs may be combined with a manual review of potential intrusions. The system can preselect probable threats and bring them to the attention of security personnel, so that they can defend customers' vehicles more effectively against threats.

We want to be able to evaluate real-time, remote IDSs for our scenario. This requires a testing environment that supports this real-time detection and emulates our remote scenario. It should simulate a complex system with interacting units that influence each other. Within clients, multiple components and sensors interact and depend on each other for their computations. Additionally, the clients' communication with the server can, to a certain extent, be unpredictable in content and sequence. It depends on implementation, connection speed, server priority and system performance. Accordingly, we require an artificial testing environment, a so-called testbed, that is based on sophisticated real-time simulations of clients from our use case. This is necessary in order to sufficiently emulate the real-world use case. Finally, we require test scenarios that subject IDSs to various detection difficulty levels.

The testbed we envision can also be used to generate historical data for off-line detection. Additionally, na\"ive data generation can be emulated by simplifying the components and simulations. Accordingly, a complete testbed can be built to fulfill our requirements while also enabling simpler scenarios.

\section{Related work}

There exist a multitude of testbeds for IDSs in various environments.
Most of these focus on different environments and use cases. An important research focus lies in testbeds for network IDSs, or NIDSs. These systems detect intrusions on different layers of Internet traffic, often focusing on the application layer. A well known example of this is {LARIAT}~\cite{rossey2002lariat}, which is based on the {DARPA} intrusion detection evaluations. It creates live traffic for protocols such as {FTP} and {SSH} and employs the NIDS as a separate node in the network that can intercept all traffic~\cite{rossey2002lariat}. \Citeauthor{shiravi2012toward} describe a more static approach, generating up-to-date datasets for multiple application protocols, including {HTTP} and {SSH}~\cite{shiravi2012toward}. These datasets can then be used for off-line detection.
Contrary to our focus, these systems aim for intrusion detection in Internet traffic. The general-purpose solutions these authors describe are too unspecific for our use case, requiring heavy adjustments or complete reworks to fit to that scenario. Our focus lies in a more specialized solution that better covers the specifics of our case.

Similarly, simulators such as ns-3~\cite{henderson2006ns3} are focused on discrete-event networks. They can be utilized when designing network protocols and interactions. This aspect can be interesting in the long run, but it does not solve the necessity of simulating the actual system behavior.

In the automotive environment, there is very limited research available. \Citeauthor{daily2016towards} discuss a testbed for heavy vehicle electronic controls~\cite{daily2016towards}. Their solution is focused on simulating sensor inputs, creating traffic for the in-vehicle network specific to commercial vehicles. \Citeauthor{huang2018atg} describe a tool that generates attack traffic for the {CAN} bus inside vehicles without making use of simulated hardware~\cite{huang2018atg}. Finally, the {HCRL} car-hacking dataset for intrusion detection~\cite{hcrl2018car} consists of real-world {CAN} traces that have been injected with attack traffic by the authors. These approaches are distinct from our scenario, as we are focused on server-side detection of intrusions into passenger vehicles.

To the best of our knowledge, there exists no work on the topic of real-time remote intrusion detection for connected vehicles.

\section{The testbed}

Our objective is to enable the comparison of various types of IDSs for the connected vehicle use case provided by our industry partner. To allow for flexible use and adaptation for different contexts, we require an artificial testing environment that fits this use case.

\subsection{Use case} 
\label{section:use-case}

\begin{figure}[htbp]
	\centering
	\includegraphics[height=30mm]{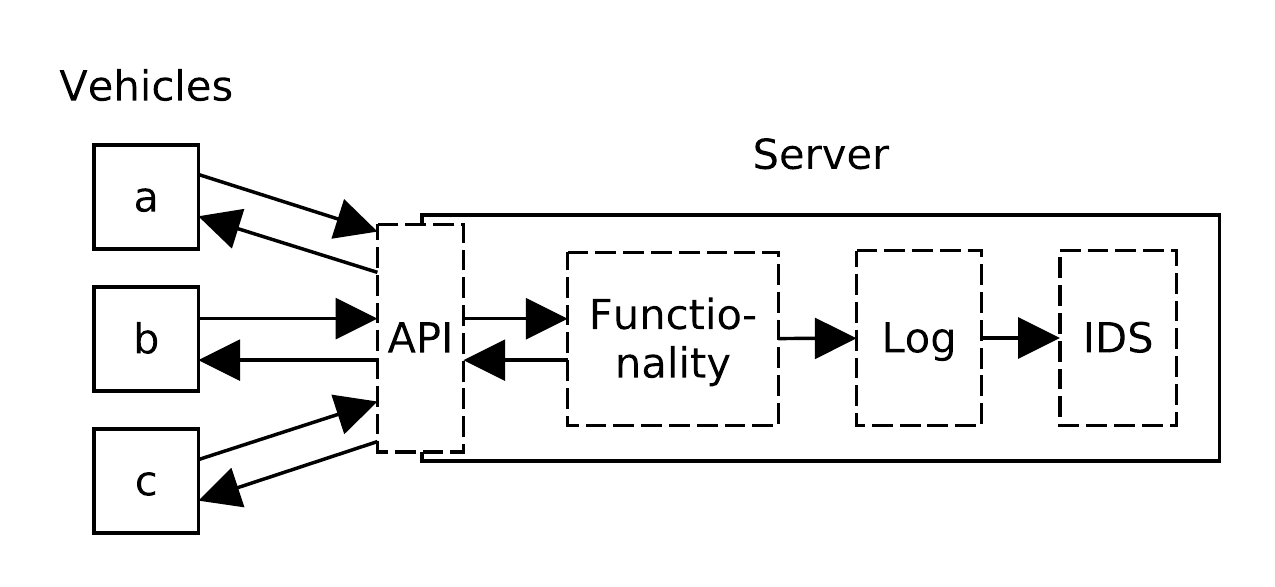}
	\caption{A model of our use case.} 
	\label{img:use-case-architecture}
\end{figure}

Our use case is based on a client-server architecture as used by our industry partner in the automotive industry. Multiple vehicles communicate with a central, manufacturer-controlled server, consisting of functional components and a logging component. Vehicle requests are handled and then forwarded to the logging component that stores information about the request in a log store (see \cref{img:use-case-architecture}). This data is used by the IDS. In the figure, arrows represent communication and components are dashed.

The server is our frame of reference, and we abstract the vehicles from an outside perspective as data providers that send requests to the server. Following similar designs by~\cite{zhang2014defending,ramadan2012intelligent,ericsson2015connected}, we assume that normal and compromised vehicles can be differentiated based on their communication with the server.
Much like the vehicles, we view the server functionality as a black box. It receives request data, processes it and sends data based on those requests to the logging component.

This component records the data that is available for intrusion detection. In our industry partner's implementation, it is server- and implementation-centric. Hence, the information stored consists of details only known to the server and its specific implementation.

\subsection{Concept}
\label{section:testbed-concept}

For comprehensive evaluation of IDSs for the real-world use case described above, we design a testbed. It is aimed at modeling that use case as closely as possible. Hence, it is simulating a client-server architecture with individual clients, corresponding to the vehicles, made up of a unique layout of different components. Each client makes use of inter-component communication and sends requests to the server. The server stub consists of a simulated logging component that generates the log data used for intrusion detection based on incoming requests from the clients. Additionally, an IDS can be added to the server for live detection. 
As we focus on intrusions in clients, the server functionality is omitted from our testbed.

\subsubsection{Clients}
\label{subsection:concept-clients}

As described before, each client consists of multiple components. These produce and consume data and are linked to a central communication unit that sends requests to the server. 
There are two types of components we can derive from real-world clients.
Simple components generate various types of data periodically, i.~e. temperature sensors. This data is then used as a basis for requests to the server.
In our model, we represent each such component with a random number generator based on a probability distribution, such as the \emph{normal distribution}. Periodically, it generates new data and sends that to the central communication unit. We refer to this component type as ``Data generator''.
More complex components in our scenario use information about the state of the client, like its position, sensor data or information about its surroundings, to allow the central communication unit to make more sophisticated requests to the server. One such example may be a request about the nearest point of interest (POI) based on the current position.
Our solution to modeling such a component is what we refer to as ``2D simulator''. 
It consists of a simulated two-dimensional environment in which an independent unit representing the client can move around.
This environment can have any defined size and different background colors. The colors represent the interpretation of certain positions by the unit moving around in the environment. A movement generator creates random movement commands for the unit. Periodically, the unit publishes information about its environment and position. This data is used both by the movement generator to adapt its movement commands to the environment as well as by the central communication unit to send requests to the server.
Finally, a central communication unit is responsible for collecting all generated data. It does only basic data transformations and then sends the result as a request to the server.

\subsubsection{Intrusions}

We define intrusions as deviations from the expected behavior of a system without the manufacturer's knowledge. That includes unauthorized access or fraud, but also software bugs (see \cref{subsection:introduction-definition}).
In our testbed, we conceptualize intrusions as modifications to individual components that lead to differences in their communication. That can mean erroneous values being generated by data generator components, or intentional bugs in the color readings of the 2D simulator.
These intrusions can be combined in various ways to form intrusion scenarios.

\subsubsection{Server}

The server we model consists of functional components and is being called by clients. Each of the requests is stored in a log for later inspection. As the server functionality is not part of our work, it is omitted. That means the testbed server consists only of the logging component. It adds some information such as the current time and an identifier to the request data, transforms it to fit our expected format and stores the result in a log.
The server can run in \emph{store mode}, meaning it stores all data that has been processed by the logging component on disk. Additionally or exclusively, it can run in \emph{detection mode}, feeding an IDS every incoming request directly, after it has been processed, for real-time detection.

\subsection{Implementation}
\label{section:testbed-implementation}

To allow for the dynamic set-up of various combinations and configurations of
clients, we use the ``Robot Operating System'' (ROS)~\cite{quigley2009ros}. This
makes defining a new set-up easy and allows for switching configurations at
will. Furthermore, while ROS is already used as a basis for automated driving
functionality (e.~g., \cite{aeberhard2015automated}).
Each client should have an individual identifier, behavior and layout, which refers to the components it contains (see \cref{subsection:concept-clients}). We use a special XML file, a so-called ROS \emph{launch file}, to define the layout of our clients. Each of its components is an individual Python program. This means we can define different layouts of components, their behavior can be individually chosen through arguments, and the clients get a unique identifier.

\subsubsection{Clients}

The clients consist of multiple components, implemented as separate programs.

The data generators representing simple components are based on \emph{NumPy}~\cite{oliphant2001scipy} random number generators. We use ten generators based on continuous probability distributions from that library.

The 2D simulator representing complex components is based on the ROS \emph{turtlesim}~\cite{faust2010turtlesim}. We reimplemented it without a graphical user interface.
Random movement commands are sent to the simulated unit. That moves according to the commands and publishes its position, speed and the current color reading.
The colors represent the interpretation of certain positions by the unit moving around in the environment.
Additionally, requests are regularly sent based on the position of the unit. They are inspired by the real-world use case of our industry partner and resemble some of the requests that clients in their scenario make. There are three request types currently implemented.
A ``Country code'' request is a simple inquiry for the associated country code for the current position of the unit: $(x, y)$. Similarly, a ``Point of interest'' (POI) request is an inquiry for a specific POI type at the current position and also contains the requested POI type: $(x, y, t)$. Finally, a ``Route'' request is an inquiry for routing to a specified target position: $(x, y, x_t,y_t) : x_t \neq x \wedge y_t \neq y$ with $(x_t, y_t)$ being the target position.
Finally, the unit has basic movement intelligence. Certain zones can be marked as ``illegal'' and the unit will try to avoid those.

\subsubsection{Intrusions}
\label{subsection:implementation-intrusions}

The reason why we simulate this environment is that we intend to evaluate a real-time remote IDS. We try to model intrusions for this purpose.
Intrusions into components can have varying difficulty levels. The rationale for these is that we aim for increasing the identifiability and with it the information entropy of the generated data compared to the expected data for easier detection difficulty levels. With this, we aim for adjusting the difficulty of detecting a compromised client.

For data generators, we currently model two types of intrusions in our system. The first is the \emph{off-value generation}.
Given the distribution interval $R := [r_{min}, r_{max}]$ containing 99.8~\% of samples of the underlying distribution function, the mean $m$ of the distribution, and its spans $s_{left} := m - r_{min}, s_{right} := r_{max} - m$. 
Further given a factor $f$ that we define.
Instead of the normal value $v \in R$, a value $v_c$ is broadcast: $v_c \in \{m - s_{left} \cdot f, m + s_{right} \cdot f\}$
Depending on the detection difficulty level, the factor $f$ differs. For level \emph{easy}, \emph{medium}, and \emph{hard}, it is $5$, $1.5$, and $1.001$, respectively. 
The smaller this factor, the higher the probability that the value could have been generated by the distribution function.

The second intrusion type, the \emph{significant error generation}, is based on a generated value from the underlying number generator. Given $m, s_{left}, s_{right}, v, f$ (see above), an erroneous value $v_e$ is broadcast. For $v \geq m$, $v_e = m + s_{right} \cdot f + v^2$, otherwise $v_e = m - (s_{left} \cdot f + v^2)$.

The 2D simulator allows for multiple types of intrusions. Firstly, the environment background has different colors. This can be changed to represent erroneous readings. An area of the simulated environment is selected and colored in a different color. The unit can pass over this area just as it normally would. In both situations the unit sends information about its position and the color it detects. Because the environment is modified, the color it reads differs, representing an intrusion.
Depending on the detection difficulty level, this area can be increased in size and its color can be modified (see \cref{table:intrusions-colors-sizes}). For different difficulty levels, we aim for varying the detectability of the erroneous color compared to the legal colors. Hence, we need a similarity measure to derive a similarity relationship between different colors

We define our colors in code as RGB (red, green, blue) with $r,g,b \in [0,255]$. The background has four different legal color options. We chose \emph{purple} (150, 140, 200), \emph{yellow} (170, 250, 140), \emph{green} (120, 180, 130) and \emph{blue} (120, 180, 200). We can imagine the colors as points in a three-dimensional space and calculate the distance between two points $p$ and $q$ using the \emph{Euclidean distance}: $ d(p, q) = \sqrt{(p_{r} - q_{r})^2 + (p_{g} - q_{g})^2 + (p_{b} - q_{b})^2} $

For our following calculations, we scale the values of the color dimensions down to $r_s,g_s,b_s \in [0,1]$. With the distance formula, we can derive the minimum and maximum distance of two points in our space: $d \in [0, \sqrt{3}]$.
The goal when choosing our legal colors was to have moderately similar colors that can still be differentiated from each other. We define that as having a distance of $d \in [0.15,0.5]$. This holds true for our four legal colors. Their distances all lie between $d_{min} \approx 0.196$ and $d_{max} \approx 0.498$. We also calculate their average distance and arrive at $d_{avg} \approx 0.256$.
We aim for increasing the distance of the erroneous color to all legal colors for easier detection difficulty levels compared to the average distance $d_{avg}$. Simultaneously, it should never fall below the maximum distance $d_{max}$ to ensure that the values are anomalous. With this requirement, we have defined a color for each difficulty level (see \cref{table:intrusions-colors-sizes}).

\begin{table}[htbp]
	\centering
	\caption{Erroneous color and its average distance to all legal colors, as well as size of erroneous area, per level.}
	\label{table:intrusions-colors-sizes}
	\begin{tabular}{ l l l l }
		\toprule
		\lheadcol{Difficulty level}        & Easy      & Medium      & Hard        \\
		\midrule
		\lheadcol{Err. color (RGB)} & 255, 0, 0 & 200, 50, 50 & 170, 80, 80 \\
		\lheadcol{Color distance}        & $1.103$   & $0.774$     & $0.590$     \\
		\midrule
		\lheadcol{Area size}              & $40~\%$   & $20~\%$     & $5~\%$      \\
		\bottomrule
	\end{tabular}
\end{table}

Secondly, the unit employs basic movement intelligence. It reacts to its surroundings and can alter its movement based on it. Certain zones can be marked as ``illegal'' and the unit will try to avoid those. This reaction can be modified: Then, the unit stays in the illegal and continuously sends its color. The detection difficulty levels for the simulated environment apply here as well. The erroneous color marks the illegal zone, and, depending on the difficulty level, its color is closer to or further from the legal colors.

Lastly, we have defined possible intrusions for the three types of requests that are sent based on the position of the unit (see \cref{table:intrusions-positional}).

\begin{table}[htbp]
	\centering
	\caption{Intrusions for positional requests.}
	\label{table:intrusions-positional}
	\begin{tabular}{ l p{62mm} }
		\toprule
		\lheadcol{Request} & \textbf{Request sent} (\textbf{N}ormal, \textbf{C}ompromised) \\
		\midrule
		C. code       & \textbf{N}: $x,y :$ current position \newline \textbf{C}: $x_c,y_c : n_x, n_y \geq 10 \wedge x_c = x \pm n_x \wedge y_c = y \pm n_y$ \\
		POI  & \textbf{N}: $t \in T :$ One of the legal types in $T$ \newline \textbf{C}: $t_c \notin T$ \\
		Route              & \textbf{N}: $x_t,y_t : x_t \neq x \wedge y_t \neq y \wedge x,y :$ current position \newline \textbf{C}: $x_{tc},y_{tc} : x_{tc} = x \wedge y_{tc} = y$ \\
		\bottomrule
	\end{tabular}
\end{table}

These types of intrusions may be detectable with domain know\-ledge-based rules. Because of their nature, we have implemented a different type of detection difficulty levels for these requests. Depending on the level, a compromised unit sends one of these compromised requests with varying probability. For levels \emph{easy}, \emph{medium}, and \emph{hard} this probability is $40~\%$, $20~\%$, and $5~\%$, respectively, making such a unit easier or harder to detect.

\section{Evaluation}

We intend to assess different aspects of our testbed to show that it fulfills its purpose, formulated as research questions (RQ).

\begin{table*}[tb]
	\centering
	\caption{Average precision and recall of a \emph{OneClassSVM} classifier on three different datasets of the respective data categories.}
	\label{table:rq-intrusion-level-effects}
	\begin{tabular}{ l l D{.}{.}{2} D{.}{.}{2} D{.}{.}{2} D{.}{.}{2} D{.}{.}{2} D{.}{.}{6} }
		\toprule
		\lheadcol{Level} & \lheadcol{Measure} & \multicolumn{1}{c}{Laplace} & \multicolumn{1}{c}{Wald} & \multicolumn{1}{c}{Color} & \multicolumn{1}{c}{POI} & \multicolumn{1}{c}{Route} & \multicolumn{1}{c}{Country Code} \\
		\midrule
		Easy & Precision (\%)    & 92.29   & 95.1     & 45.4     & 72.80  & 74.42   & 59.25 \\
		& Recall (\%)       & 100.0   & 100.0    & 8.26     & 100.0  & 100.0   & 14.18 \\
		Hard & Precision (\%)    & 91.98   & 87.93    & 1.99     & 74.0   & 73.66   & 73.64 \\
		& Recall (\%)       & 100.0   & 44.69    & 0.19     & 100.0  & 100.0   & 30.12 \\
		\midrule
		Difference & Precision ($pt.$) & -0.31 & -7.17  & -43.41 & +1.2 & -0.76 & +14.39 \\
		& Recall ($pt.$)    & \multicolumn{1}{l}{\phantom{...}$\pm$0}  & -55.31 & -8.07  & \multicolumn{1}{l}{\phantom{l}$\pm$0} & \multicolumn{1}{l}{\phantom{`}$\pm$0}  & +15.94 \\
		\bottomrule
	\end{tabular}
\end{table*}

\researchquestion{Do we solve the problems of existing datasets?}

In our research of IDSs, we identified three problems in the datasets used for evaluation.

\slimparagraphy{Too few entries}
Many IDS techniques require a significant amount of data for sufficient evaluation which can be difficult to obtain~\cite{cannady1998artificial, knorr2000distance}.
Our testbed can arbitrarily generate new entries, meaning the appropriate number of entries can be generated as needed.

\slimparagraphy{Class imbalance}
Training data showing class imbalance is an important problem if it is used for machine learning systems, as it leads to worse classifier performance~\cite{chawla2004special}.
We can tune the testbed, but we cannot precisely predict the resulting class imbalance in the data, so we have to evaluate it in the following.

\slimparagraphy{Redundancy}
Data redundancy leads to a bias in \emph{ma\-chine-learn\-ing} systems~\cite{li2012efficient, sindhu2012decision}.
Mitigating this by deleting the redundant records leads to fewer entries in the dataset, an issue in itself.
The redundancy in our generated data can also not be predicted for every scenario, so we evaluate it in the following.

From these problems, we can derive quality metrics. As we consider the problem of too few entries solved with the testbed, we evaluate the class imbalance and redundancy of the generated data.

As a quality measure for the class balance of a dataset we use the \emph{dispersion index}, defined as $\frac{\text{variance}}{\text{mean}}$. This measure is $0$ for an exactly balanced dataset.
To measure redundancy, we count the duplicates in our datasets as follows: Consider a log line $l_i$ in line $i$ of a log file. We define this log line as a duplicate if the following holds: $i > 1 \wedge \exists$ $l_{i-1}$ where the unit's identifier, its position, the data sent and the label in $l_i$ and $l_{i-1}$ are identical, then $l_i$ is a duplicate.

\expsetupheader{}
We generated a dataset containing 10 million data points and analyzed it. The dataset contains data points for 14 data categories with approximately 715,000 data points each. We consider each kind of data generated by a component of the testbed as a separate data category, allowing for more precise evaluation. That means we calculate the measures for each data category separately. For a more detailed explanation of these categories, see \cref{section:testbed-implementation}. Regarding the aspect of class imbalance, we consider two classes, the normal and the intrusion class. For each data category, we calculate the relative class imbalance. We consider relative deviations of less than $1~\%$ (7,150 data points) as ideal, and of less than $5~\%$ (35,750 data points) as good. That corresponds to a dispersion index of $18$ or less for ideal, and of $450$ or less for good results.
Regarding duplicates in the data, we consider $0~\%$ to be ideal and anything less than $1~\%$ to be good.

\resultsheader{}
We split up the results based on the underlying component. All data generators behave identically regarding intrusions, so their results are grouped as \emph{Generators}. The same is true for positional requests, grouped as \emph{Positional}. The results for color requests are listed separately (see \cref{table:datasets-solves-problems}).
All data categories succeed in meeting our strict duplicate targets. The class balance is not perfect, but for color requests we observe good and for positional requests ideal results. The \emph{Generators} show an acceptable dispersion index below our threshold of $450$.

\begin{table}[htbp]
	\centering
	\caption{The results for different categories of data, grouped.}
	\label{table:datasets-solves-problems}
	\begin{tabular}{ l l l l }
		\toprule
		\lheadcol{Measure} & \lheadcol{Generators} & \lheadcol{Positional} & \lheadcol{Color} \\
		\midrule
		Number of elements     & 7,021,192   & 2,160,908   & 817,900 \\
		Class dispersion index & $250.8$     & $7.8$       & $32.1$ \\
		Duplicates             & $0~\%$      & $0.016~\%$  & $0.009~\%$ \\
		\bottomrule
	\end{tabular}
\end{table}

\researchquestion{Can we show the effect of varying detection difficulty levels?}

We implemented detection difficulty levels aiming at subjecting IDSs to different difficulty levels. Data generated with difficulty level \emph{hard} is supposed to lead to reduced detection accuracy compared to \emph{easy}, as the intruded samples for \emph{hard} lie much closer to the normal data. To evaluate this, we created an IDS prototype based on machine learning. For this research question, we need to introduce some additional prerequisites.

\slimparagraphy{Handling heterogeneous data}\label{subsection:prerequisites-heterogeneous-data}
As we have described before, the generated data varies between different routines in the logging component. Each of these routines represents some server functionality, and in our real-world use case these require or produce different data. 
We want to use the variability in the data generated by the testbed to show distinct effects. Hence, we score systems individually for each type of data in our experiments.

\slimparagraphy{Libraries used}\label{subsection:prerequisites-libraries-used}
We make use of Python and machine learning, with our library of choice being \emph{scikit-learn}~\cite{pedregosa2011scikit}. Apart from classification, we utilize included pre-processing and metrics modules. From this library, we employ the \emph{OneClassSVM} classifier, which is based on the \emph{libsvm}~\cite{chang2011libsvm} library for computations.

\slimparagraphy{Definition of positives}
We regard those data points as positives that are intrusions.
This interpretation is common in intrusion detection research (see e.g. \cite{kim2007immune}).

As our IDS prototype does not consider client-specific profiles, we do not expect the detection difficulty levels for categorical data in the form of positional requests to lead to any differences. 
Contrary to that, difficulty levels implemented for the data generators and color values focus more on \emph{anomaly-based} detection. They are expected to have an effect on the outlier detection algorithm that we apply here.
If we find the difference of precision or recall between the difficulty levels to be over a certain threshold, we assume changing the level leads to significant differences in detection accuracy. Our threshold for this is $5$ percentage points.

\expsetupheader{}
We generated 1 million data points for detection difficulty levels \emph{easy} and \emph{hard} respectively, and sampled three sets of 100,000 data points for both levels.
\emph{OneClassSVM} classifiers with default parameters were trained for each level, data category and sample in a total of $84$ rounds.

\resultsheader{}
After three iterations with three different datasets, we arrived at an average precision and recall percentage of the classifiers for each data category and both difficulty levels (see \cref{table:rq-intrusion-level-effects}).
As expected, precision and recall do not change significantly for the positional requests \emph{POI} and \emph{Route}. We cannot explain the difference observed for \emph{Country code} data. For the data generators \emph{Pareto}, \emph{Rayleigh}, \emph{Uniform}, \emph{Wald}, and \emph{Weibull} we find a successful increase in difficulty, shown exemplary in the \emph{Wald} column in \cref{table:rq-intrusion-level-effects}. For \emph{Gaussian}, \emph{Gumbel}, \emph{Laplace}, \emph{Logistic} and \emph{von Mises} data, we find no significant difference in precision and recall between detection difficulty levels. These results are shown exemplarily in the \emph{Laplace} column of \cref{table:rq-intrusion-level-effects}. The other data generator categories are omitted from the table for brevity. Finally, \emph{Color} data shows a visible effect of increasing the difficulty level.

In conclusion, additional and more specific experiments are necessary to judge the quality of individual difficulty levels for different data categories. The currently implemented broad intrusions seem to have different effects for different data categories, showing the intended effect for most, but not all types of requests.

\researchquestion{Can we reproduce data of similar distribution?}

Data generated by a testbed is often not considered static but needs to be reproduced multiple times. For comparable behavior over multiple runs of the same set-up, it is essential that the generated data is similarly distributed.

For our data generators, we consider this to be the case. The underlying number generators produce data based on a statistical distribution, resulting in equally distributed data over multiple runs. Exemplary, we show this for our \emph{normal distribution}-based generator. The more complex component is the 2D simulator. For each of the request types, we want to assess if the distribution of their possible forms stays similar over multiple runs.

For evaluation, we generate three datasets. To minimize the influence of sampling errors on the result, each set is generated with 250,000 data points. We consider the first set the baseline, and the following sets are compared to it. We give the coefficient of determination \rtwo{} for both sets compared to the baseline. If the baseline distribution is perfectly reproduced by the following set, this measure is $1$. We use the \rtwo{} as it is commonly used to show how well data points correspond to a given baseline.

\slimparagraphy{Data generators}
\label{subsubsection:reproducibility-data-generators}
First, we want to show the relative distribution of data generated by a \emph{normal distribution} generator, exemplary for our data generators.
We split the data up in bins of size $0.1$. Then, we calculated the relative frequency of data points in each bin and compared the sets. For data points of the normal class, the \rtwo{} value of Set 2 compared to the baseline set is approximately $0.9995$, for Set 3 it is approximately $0.9996$. For the intrusion class, the \rtwo{} value of Set 2 compared to the baseline set is approximately $0.9998$, for Set 3 it is approximately $0.9997$.
As expected, this component generates sufficiently reproducible data with almost ideal \rtwo{} values.

\pgfplotsset{colormap={BW}{rgb255=(44,44,44) color=(white)}}
\pgfplotsset{
	coordinatematrix/.style={
		width=50mm,
		height=50mm,
		xlabel=\empty,
		ylabel=\empty,
		xtick=\empty,
		ytick=\empty,
		xmin=-0.5,xmax=24.5,
		ymin=-0.5,ymax=24.5,
		enlargelimits=false
	}
}

\begin{figure}[htbp]
	\parbox{.32\linewidth}{
		\centering
		\includegraphics[width=\linewidth]{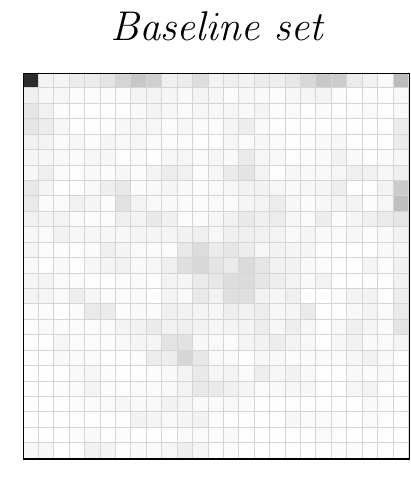}
	}
	\hfill
	\parbox{.32\linewidth}{
		\centering
		\includegraphics[width=\linewidth]{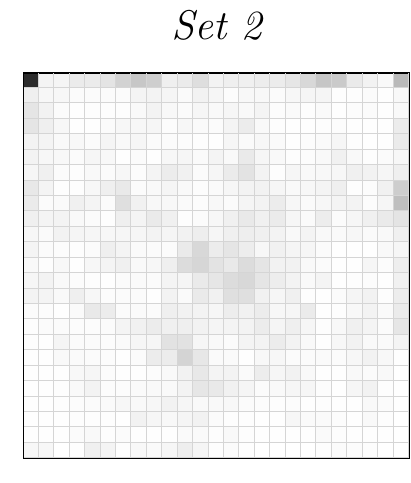}
	}
	\hfill
	\parbox{.32\linewidth}{
		\centering
		\includegraphics[width=\linewidth]{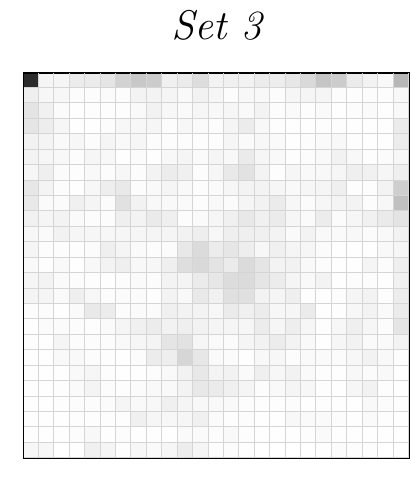}
	}
	\caption{Relative frequency of a coordinate bin in the simulated environment.} 
	\label{plot:distribution-positions}
\end{figure}

\slimparagraphy{2D simulator positions}
Next, we evaluate the relative frequency of the simulated unit being located at various positions of the simulation. This is tricky to visualize, as we have a significant number of points to compare. We approach this problem with heat maps. For easier visualization, we divide the coordinate space into bins of $20 \times 20$ pixels in size. The relative frequency of each bin is signified by the shade of the cell that represents it (see \cref{plot:distribution-positions}). For higher frequencies, the box shifts to a darker shade. The highest frequency is visualized in black.
The heat maps all show a similar pattern. Most noticeable is the dark cell in the top left corner. The simulated unit seems to disproportionally often remain in that area. Along the edges and around the center we see more areas of higher relative frequency, all of which are mirrored in the other sets.
This intuition is confirmed when calculating the \rtwo{} value. For Set 2 compared to the baseline set it is approximately $0.9965$, for Set 3 it is $0.9962$.

\slimparagraphy{2D simulator requests}
Finally, we consider the requests made by the 2D simulator component, which consist of categorical data. We will discuss our evaluation in detail for point of interest (POI) pairs and list only the results for the remaining data categories.

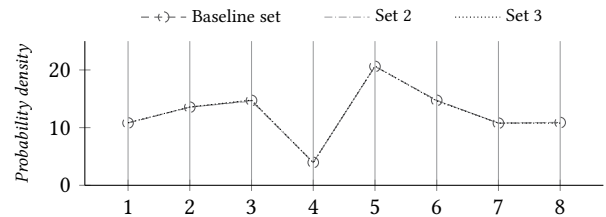
\begin{figure}[htbp]
	\centering
	\begin{tikzpicture}
	\begin{axis}[
		scalabilityplot,reproducibilityplot,histogramline,
		width=\linewidth,
		symbolic x coords={1, 2, 3, 4, 5, 6, 7, 8},
		ymin=0, ymax=25]

		\addplot+[sharp plot,mark=o]
		coordinates {
			(1, 10.81) (2, 13.54) (3, 14.71) (4, 3.96) (5, 20.61) (6, 14.73) (7, 10.77) (8, 10.88)
		};

		\addplot+[sharp plot]
		coordinates {
			(1, 10.84) (2, 13.55) (3, 14.58) (4, 3.96) (5, 20.63) (6, 14.71) (7, 10.8) (8, 10.92)
		};

		\addplot+[sharp plot]
		coordinates {
			(1, 10.85) (2, 13.58) (3, 14.78) (4, 3.93) (5, 20.67) (6, 14.62) (7, 10.79) (8, 10.78)
		};

		\legend{Baseline set, Set 2, Set 3}

		\end{axis}
	\end{tikzpicture}
	\caption{Relative frequency of POI pairs.} 
	\label{plot:distribution-poi-pairs}
\end{figure}

POI pairs consist of a POI type and a POI result. Clients request a POI of a specific type for their location. The server retrieves the POI result and stores the requested type and the result as a log entry.
There are two legal POI types with three possible results each, and two illegal types that both map to the same result, namely ``Invalid''. Each allowed combination of a POI type and result is one possible form of a POI request. In \cref{plot:distribution-poi-pairs} we compare the relative frequency of possible forms of POI pairs in the three sets. We see almost identical distributions for different sets. This is confirmed by the \rtwo{} value. Comparing Set 2 to the baseline set the \rtwo{} is approximately $0.9995$, for Set 3 it is approximately $0.9998$.

We find similar results for the other request types. For ``Color'' requests, the \rtwo{} value of Set 2 compared to the baseline set is approximately $0.9999$, for Set 3 it is approximately $0.9997$. The categories ``Country code'' and ``Route'' have an almost ideal \rtwo{} value of approximately $0.9999$ for both Set 2 and Set 3 compared to the baseline set.

\slimparagraphy{Conclusion}
When generating data multiple times, the resulting distributions of values are ideally the same or very similar. This objective is fulfilled for all testbed components, including the more complex movement patterns of the 2D simulator.

\researchquestion{Can we evaluate various types of IDS?}

There are various, heterogeneous IDS approaches. Existing evaluation options are often limited to specific types of systems. For example, authors utilizing \emph{anomaly-based} detection can use general-purpose datasets such as the \emph{Iris flower set}~\cite{duda1973pattern} for evaluation. For \emph{signature-based} systems, such datasets offer no sensible evaluation option as they are missing intrusions that could be detected. We want to show that our testbed theoretically allows the evaluation of various types of IDSs.

\slimparagraphy{Anomaly-based detection}
IDSs utilizing \emph{anomaly-based} detection require data with some form of anomalous behavior that they can detect. This use case is covered by our data generators. They are based on probability distributions, which allow us to define normal behavior as that which is most likely to happen based on the underlying distribution. Accordingly, we introduced intrusions that aim at being discernible from normal data (see \cref{subsection:implementation-intrusions}).

\slimparagraphy{Signature-based detection}
To evaluate \emph{signature-based} detection systems, our data requires well-defined signatures that can be coded into these systems. Our 2D simulator component aims at providing these. The data it generates is used to create requests aimed at emulating domain knowledge-based patterns. Similarly, we introduced intrusions that are based on breaking some defined rule. Currently, they are built to also be detectable by \emph{anomaly-based} systems.

\slimparagraphy{Advanced detection systems}
Advanced systems such as \emph{Artificial Neural Networks} (ANN) or \emph{model-based reasoning} approaches are currently impossible to evaluate with our testbed. We have not defined long-term behavioral patterns or specific user profiles that would allow for that. These systems may best be evaluated with real-world data.

\slimparagraphy{Conclusion}
We consider the testbed to be sufficient for evaluating \emph{anomaly-based} as well as \emph{signature-based} systems on a theoretical basis. Due to the automatic labeling of testbed data, the detection accuracy can be measured effectively. For more advanced detection systems, our testbed does not offer adequate options for evaluation.

\researchquestion{How well does the testbed scale?}
\label{rq:scalability}

We are interested in the scalability of our testbed regarding the system load when increasing the number of simulated components. Ideally, it scales almost linearly, as this allows running arbitrarily complex simulations if necessary. We regard the metrics \emph{Start-up time}, \emph{CPU load}, and \emph{Main memory usage}. We perform a \emph{regression analysis} to approximate the time complexity based on our measurements. For each analysis, we give the \emph{mean squared error} \mse{} and \emph{coefficient of determination} \rtwo{} to assess the quality of the estimation compared to the measurements. For better estimations, the \mse{} lowers towards 0 and the \rtwo{} increases towards 1.

For our experiments, we used a machine with a 12-core 2.5 GHz \emph{QEMU}~\cite{bellard2018qemu} virtual CPU and 32 GB of main memory.
Because we could only measure reliably for component counts of up to $500$, we do not generalize further. For component counts larger than $500$, we would need more measurements to be able to sufficiently predict the limiting behavior.

The measurements are split up in cycles and rounds. Each cycle is identified by the number of components that are started. Each round represents one measurement. We measured repeatedly and took the average for our results.
For the start-up time, the maximum standard deviation of all measurements was $0.26$ seconds, or $1.3~\%$, with an average value of $0.07$ seconds.
For the CPU load measurements, we calculated a maximum standard deviation of $0.18$ points, or $1.2~\%$, with an average value of $0.4$ points.
The maximum for the main memory measurements was $0.59$ percentage points, or $1.58~\%$, with an average value of $0.08$ points.
Hence, we consider the measurements to be sufficiently reproducible.

\slimparagraphy{Results}
The measurements for start-up time, CPU load, as well as main memory usage grow linearly, so we estimate them with linear regression. We will describe our approach in detail for the main memory usage, and only list the results for the other measures.
After multiple iterations with subsets of our measurements, we obtain for the main memory usage : $g_m(n) \approx 0.1491n + 0.447$ with $\operatorname{MSE} < 1.262$ and $R^2 > 0.997$ for the complete set of measurements (see \cref{plot:estimated-function-main-memory-usage}). In the figure, the dots are the mean, with the bars above and below indicating the maximum and minimum measurement.

\newcommand{\errorbarhint}{Each dot is the mean, with the bars above and below each dot indicating the maximum and minimum measurement.}

\pgfplotsset{
	rqmainmemplot/.style={
		xmin=0, xmax=525,
		ymin=0, ymax=80
	}
}

\begin{figure}[htbp]
	\centering
	\begin{tikzpicture}
	\begin{axis}[
	scalabilityplot,rqmainmemplot,
	xtick distance=100,
	minor x tick num=1,
	ylabel=\textit{Memory usage in \%}
	]

	\addplot[smooth,no markers,draw=TumDarkGrey,fill=none,domain=0:525]
	{0.44663083786 + 0.149137883989 * x}; \label{gmplot}

	\end{axis}

	\begin{axis}[
	scalabilityplot,rqmainmemplot,
	axis x line=none,
	axis y line=none,
	xlabel=\empty,
	ylabel=\empty
	]

	\addplot+ [errorbarplotstyle]
	coordinates {
		(2,0.59) += (0,0.0) -= (0,0.01)
		(50,8.075) += (0,0.515) -= (0,0.765)
		(100,15.67) += (0,0.91) -= (0,2.18)
		(150,22.925) += (0,1.925) -= (0,1.725)
		(200,30.415) += (0,2.095) -= (0,1.985)
		(250,37.63) += (0,2.29) -= (0,1.36)
		(300,45.115) += (0,2.415) -= (0,2.115)
		(350,52.44) += (0,2.47) -= (0,2.77)
		(400,60.51) += (0,1.86) -= (0,2.55)
		(450,67.175) += (0,1.705) -= (0,2.015)
		(500,75.95) += (0,1.7) -= (0,2.8)
	}; \addlegendentry{$mem$}

	\addlegendimage{/pgfplots/refstyle=gmplot}\addlegendentry{$g_m(n)$}

	\end{axis}

	\end{tikzpicture}
	\caption{Main memory usage and approximation $g_m(n)$.} 
	\label{plot:estimated-function-main-memory-usage}
\end{figure}
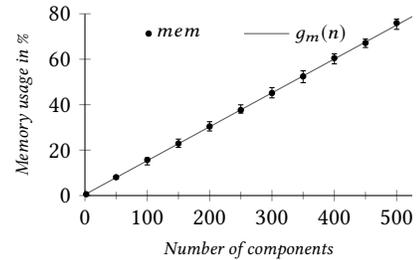

From our approximation, we can derive the limiting behavior: $f_m(n) = \mathcal{O}(0.1491n + 0.447) = \mathcal{O}(n), n \to 500$.
Accordingly for the start-up time, we obtain: $g_s(n) \approx 0.0476n + 1.141$ with $\operatorname{MSE} < 0.143$ and $R^2 > 0.998$. We conclude: $f_s(n) = \mathcal{O}(0.0476n + 1.141) = \mathcal{O}(n), n \to 500$. We assume that the constant $1.141$ represents the cold-start time.
For the CPU load, we obtain: $g_c(n) \approx 0.0127n + 9.756$ with $\operatorname{MSE} < 0.031$ and $R^2 > 0.988$. We conclude: $f_c(n) = \mathcal{O}(0.0127n + 9.756) = \mathcal{O}(n), n \to 500$.

\slimparagraphy{Conclusion}
We can show for all our measures that a linear approximation can sufficiently explain our measurements. We derive a limiting behavior of $\mathcal{O}(n)$ for component counts of up to $500$, meaning our testbed scales linearly. For a valid estimation for higher $n$, we would need additional measurements.

\section{Limitations}

Our testbed concept and implementation were created from the ground up. To be able to simulate our use case, we had to simplify it and create exemplary implementations based on a more basic concept of the use case. This abstraction means that the system by itself is not representative for the real world. To reduce this risk, we closely modeled the scenario after the real-world use case, with the testbed generating data that resembles the actual data found in the system employed today.
Using synthetic data is delicate and not the perfect solution. It is a necessary compromise though, as real-world data cannot be accessed in relevant quantities. Other authors have made this compromise for the same reason~\cite{huang2018atg}. When using data from the testbed it is important to keep those risks in mind.
Finally, the IDS we used for evaluation is only a simplified machine learning based implementation. This does not pose a threat to the validity of our results, as we only measured differences in detection difficulty, not absolute ease of detection.

\section{Conclusion and future work}

More and more vehicles are connected, exposing them to the risk of attacks. IDSs can be employed as a countermeasure. To be able to evaluate different IDSs, we conceptualized and implemented a testbed. It enables the comparison of various IDSs in a real-world scenario. The testbed is adapted to our automotive use case but can be quickly adjusted to numerous other scenarios.

In the evaluation, we find that the testbed fulfills its objectives.
Its architecture and modularity allow for various configurations and the evaluation of real-time, remote IDSs.
To assess the fitness for purpose of the testbed, we evaluate multiple quality metrics derived from related works. We find that limitations of existing datasets used in research are solved by our implementation. When generating data repeatedly, the testbed exhibits high performance and reliable behavior. This allows the reproduction of equally distributed data, which is beneficial for repeated evaluation and cross-validation. Additionally, we find that the testbed can be used to evaluate the various types of IDSs we have identified in literature.
We also assess the quality of the data generated by the testbed. The detection difficulty levels we have introduced lead to measurable differences in detection accuracy, allowing for varied evaluation of IDSs. This confirms that IDSs are challenged when trying to detect intrusions in the data generated by our testbed.
To conclude our evaluation, we evaluate the performance of the testbed. We find that with additional components the testbed scales linearly. This allows for complex simulations to be carried out.
We have discussed some limitations to our solution above. This mainly regards the number of different components and the generated data.

Our testbed is a first step towards a comprehensive solution for comparing IDSs. More advanced behavior patterns can be implemented, including multiple units interacting in the 2D simulator component within a shared environment.
Additionally, more subtle intrusions may be implemented based on domain knowledge.
Still, the current state of the testbed is sufficient to evaluate most types of IDSs employed today while offering multiple improvements over existing solutions.

In conclusion, our testbed allows the flexible and reliable evaluation of IDSs of various types and in multiple scenarios. Common problems of existing evaluation approaches are solved, which allows a more up-to-date and effective comparison of different IDSs. Because of the reproducibility of the generated data, the same IDS can be tested multiple times to ensure consistent performance. Additionally, the generation of arbitrary amounts of new test data allows for effective cross-validation. Finally, other researchers can create custom evaluation scenarios with the testbed, without having to rely on outdated or fixed datasets.
The source code of our testbed is available for inspection, modification and use\footnote{\href{https://github.com/tum-i22/rritbed}{\texttt{https://github.com/tum-i22/rritbed}}}.

\begin{acks}
	This work was supported by the Deutsche Forschungsgemeinschaft (DFG) under grant no. PR1266/3-1, Design Paradigms for Societal-Scale Cyber-Physical Systems, and the Federal Ministry of Education and Research (BMBF) under grant no. 5091121.
\end{acks}

\bibliographystyle{ACM-Reference-Format}
\bibliography{references}


\end{document}